# A PHASE TRANSITION IN A SIMPLE PLASMA MODEL*

I.L. Iosilevskii

*Moscow Institute of Physics & Technology (MIPT)*

A phase transition of gas-liquid type with an upper critical point is examined which arises in a model of charges of one sign on compensating background (OCP). The phase transition parameters are dependent on the detailed assumptions about the compressibility of the background, but the occurrence of this transition is independent on the background equation of state. In the electron-gas model ("*jellium*"), this transition appears to rule out Wigner crystallization. A variational principle in statistical mechanics is used to derive so-called Double-OCP model for a superposition of two one-component plasma models for charges of opposite sign. The free energy of this model sets an upper bound to that of a real plasma. Situations are discussed where this transition should manifest itself in anomalies in the approximate description of a non-ideal plasma.

A model for a one-component plasma (OCP) consists of a system of charges of one sign against a passive or compensating background of charge of the opposite sign, and this plays an important part in plasma theory. In addition to the traditional applications in metal physics and astrophysics (see review [1-3]), the OCP model is of increasing importance in the description of more complex plasma models [4-6]. The OCP model is exceptionally important because of its simplicity of method in checking the theoretical conclusions on real plasmas [7-10]. However, OCP is not simple in essence although it is the simplest of the plasma models in a computational respect. For example, the proof of a thermodynamic limit in the OCP model [10] is more complicated than that for plasma containing differing charges [11]. OCP also shows negative pressure and compressibility over a considerable parameter range [1, 12], which so far has not received an unambiguous interpretation [7, 13-17]. One can retain the positive signs for these quantities by defining them as in [15-17], but this does not eliminate the question of the thermodynamic stability or the need to reconsider all relationships involving volume variation.

These difficulties are related to unclear definition of the properties and thermodynamic role of the background in the OCP model, which in general directly concerns the thermodynamic stability, as well as the phase transition type and position. In the original form, which was free from ambiguity, the OCP model implies the presence of an incompressible background. This means that the system volume is fixed, and quantities related to volume variations are not defined. The model cannot contract spontaneously, while the negative values for the corresponding formal expressions for the pressure and compressibility [1, 12] do not mean loss of thermodynamic stability. The only phase transition identified in the OCP model (crystallization) involves no change of volume, only a structure rearrangement [#].

A more realistic OCP model is that where the compensating background remains homogeneous but has nonzero compressibility. In that form, crystallization is accompanied by volume change (see estimates [20]). However, a more important consequence is that there is an additional phase transition, which may be termed of the gas-liquid type, and whose parameters are substantially dependent on the detailed assumptions about the background, although it is notable that its existence may be established without reference to any approximate calculations.

______________________________

(*) The evidence in this paper was reported at the 2nd Soviet Conference on Equations of State in Cheget in 1980 and at the 6th Conference on Low-Temperature Plasma Physics in Leningrad in 1983.

(#) An example of a system where spontaneous contraction is artificially hindered is one examined by Monte-Carlo methods for a {N,V,T} ensemble of hard spheres with additional 'square-well' attraction [18] (see also [19]), which records negative pressures and compressibility.

______________________________

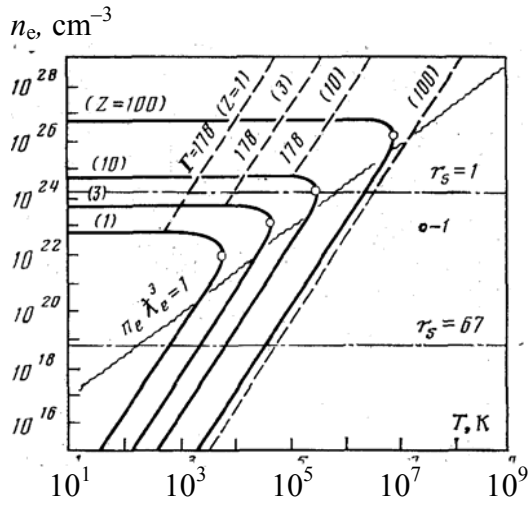 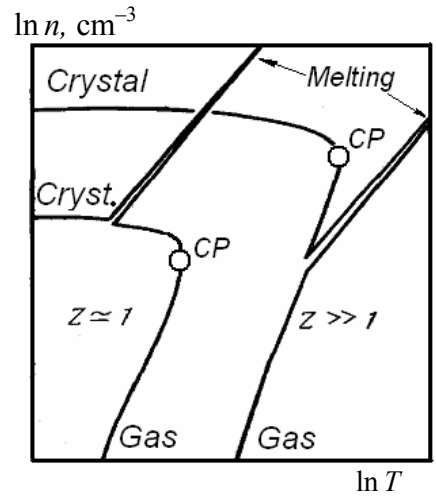

**Fig. 1**. OCP of classical ions $Ze$ on the uniformly compressible background of ideal electron Fermi-gas (*1* - the critical point; $r_S$ – Brackner parameter).
**Fig. 2**. General form of the phase diagram for classical-charge OCP on uniformly compressible background of electron ideal Fermi-gas.

**Classical OCP model**. The system studied in most detail is that of classical point particles of charge $q \equiv Ze$ on a 'rigid' (incompressible) compensating background. This has been examined by asymptotic methods (see review [7-9]) and by Monte-Carlo methods [1, 12, 20] and by methods of molecular dynamics [21]; it has now been examined in detail virtually throughout the parameter range. The equation of state for this model is a function of a single dimensionless parameter $\Gamma$ (or the equivalent $\Gamma_D$)

$$\Gamma \equiv q^2(4\pi n/3)^{1/2}/kT \equiv (\Gamma_D^2/3)^{1/3}, \qquad \{\Gamma_D \equiv q^2/kTr_D \equiv (4\pi n)^{1/2}(q^2/kT)^{3/2}\} \tag{1}$$

where $n$ is the charged-particle concentration, $T$ - temperature, $r_D$ - Debye radius, and $k$ is Boltzmann's constant.

Monte Carlo methods have indicated a unique phase transition in this model: crystallization, whose position in the density-temperature phase diagram corresponds to the line $\Gamma = const \approx 155$ [12, 20] ($\Gamma_D \approx 3340$)(*).

We now consider the classical OCP model, where the background is provided by a hypothetical ideal electron gas, with the electrons not correlated with the main OCP charges. The system as a whole is assumed to be electrically neutral, $n_e = Zn$ ($q = Ze$, where e is the electron charge). The overall thermodynamic functions in the model are made up of the corresponding charge and background functions. We assume that the background can contract spontaneously as a whole, but at the same time it cannot contract (correlate) around each ion separately, so all the individual screening of ions is due to correlation between the ions only.

Any interacting system obeys the Gibbs - Bogolyubov inequality (see, for example, [23]), in accordance with which the dependence of the thermodynamic quantities on the coupling constant $\lambda$ characterizing the interaction (here $\lambda \equiv e^2$) should be sufficiently strong. In particular, the free energy $F$ as a function of $\lambda$ should be convex for any $\lambda$ [15]:

$$(\partial^2 F/\partial \lambda^2) \leq 0 \tag{2}$$

At this is a one-parameter model, (2) leads to a constraint on the dependence of the thermodynamic quantities on $\Gamma$. As a result, the interaction energy U should increase with $\Gamma$ not less than $\Gamma \cdot const$ [24](#) and the same applies to the non-ideality correction in the equation of state:

$$d\ln|U/NkT|/d\ln\Gamma = d\ln|\Delta P/nkT|/d\ln\Gamma \geq 1 \tag{3}$$

---
(*) According to a recent calculation, $\Gamma \approx 178 \pm 1$ [22].
(#) In the Debye approximation in Grand Canonical Ensemble [25, 26], commonly used in non-ideal plasma theory, inequality (3) is violated for $\Gamma \geq 0.78$ [26].

**Table 1.** Critical-point parameters (**I**) and spinodal parameters for $\Gamma \to 0$ (**II**) for a model of classical charges $Ze$ on a compensating background of an ideal electron Fermi-gas.

|   | $Z$ | 1 | 2 | 3 | 10 | 30 | 100 | 1000 |
|---|---|---|---|---|---|---|---|---|
| I | $kT_{cr}$ (Ry/particle) | 0,535 эB | 0,152 | 0,314 | 2,17 | 10,6 | 55,8 | 1232 |
|   | $\Gamma_{cr}$ | 8,86 | 12,6 | 16,3 | 42,7 | 117 | 376 | 3690 |
|   | $(r_S)_{cr}$ | 5,73 | 3,32 | 2,43 | 1,00 | 0,467 | 0,206 | 0,044 |
|   | $(n_e\lambda_e^3)_{cr}$ | 7,22 | 4,90 | 4,22 | 3,30 | 3,03 | 2,94 | 2,89 |
|   | $\{p/(n+n_e)kT\}_{cr}$ | 0,125 | 0,129 | 0,130 | 0,130 | 0,129 | 0,128 | 0,127 |
| II | $(r_S)_{spinodal\ (liquid)}$ | 3,08 | 1,94 | 1,48 | 0,664 | 0,319 | 0,143 | 0,031 |
|   | $\Gamma_{spinodal\ (vapor)}$ | 5,76 | 8,39 | 11,0 | 28,9 | 79,6 | 256 | 2518 |

Here $p$ is the pressure and $N$ is the total number of charges.

Therefore, when there is isothermal compression, the correction in the equation of state $\Delta p$ is negative and should increase in magnitude not more slowly than as $n^{4/3}$. If we use an ideal Boltzmann gas as the background in the model (i.e., $p_{id} = n_e kT$), the system of charges plus background sooner or later becomes absolutely unstable at any temperature and for any degree of dilution with the background ideal gas. This means that an OCP with such a background does not exist in thermodynamic equilibrium (collapses) for any parameters, while the equation of state obtained for $\Gamma \le 3$ with positive compressibility [1] corresponds to an unstable (metastable) phase.

**OCP with quantum effects**. The above result agrees with Lenard and Dyson's assertion [27] (see also [28]) that if thermodynamic equilibrium is to exist in a system of charges differing in sign, it is necessary for at least one of the types of particle to obey Fermi statistics. In the OCP model, the pressure will increase as $n^{5/3}$ for $n \to \infty$, which is sufficient for stability, since together with the constraint of (3) there exists a lower simple and effective bound [10](*)

$$U/(NkT) = 3\Delta P/(nkT) \ge -0,9 \cdot \Gamma \sim n^{1/3}. \tag{4}$$

Therefore, in an OCP model in which either the charges, or the background, or the two together obey Fermi statistics one gets stability at the two limits $n \to 0$ and $n \to \infty$. Nevertheless, the gas-liquid phase transition persists in this form of model, but now with the critical point. In fact, the relation between the $\Gamma$ of (1) and the degeneracy parameter $n_e\Lambda_e^3$ ($\Lambda_e$ is the de Broglie length) is such (Fig. 1) that the effects of the strong Coulomb interaction on isothermal compression always occur before the plasma becomes degenerate at low temperatures or for sufficiently large $Z$ ($Z^4 Ry/kT \gg 1$). Consequently, everything said above about the inevitable stability loss in OCP with ideal gas background still applies.

Consider a system of classical point particles having charge $Ze$ against the background of an ideal electron Fermi gas. The equations of state for the two components are known, so the parameters of this phase transition can be calculated accurately. Figure 1 and Table 1 give results for various values of the charge $Z$: the critical-point parameters and those of both spinodals (i.e., the bounds of absolute thermodynamic instability). In these calculations, an interpolation expression was used for the equation of state for an ideal electron gas as proposed in [29], while that from [12] was used for the charge subsystem. In the latter case, no allowance was made for the difference between gas branch of the equation of state [12] and the crystalline one [20] because they are similar on calculating the spinodal parameters for $\Gamma > 178$.

We now explain the results. Figure 1 shows that the critical point lies near the degeneracy boundary for the background for all $Z$: the line $n_e\Lambda_e^3 = 1$. For all $Z$, the critical temperature certainly exceeds the real value for gas-liquid transition, which is a direct consequence of the artificial inhibition of individual screening for each charge by the background (or charges) of the opposite sign, which is characteristic of this OCP model.

---

(*) The following stronger bound is assumed to be valid for OCP, although it has not been strictly proved [28]:

$$U/NkT \ge (U/NkT)_{Crystal} = -0,89593\ \Gamma.$$

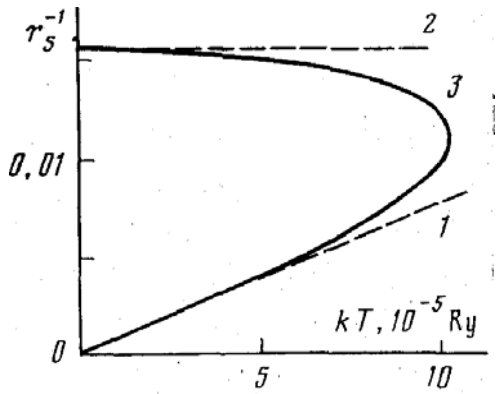 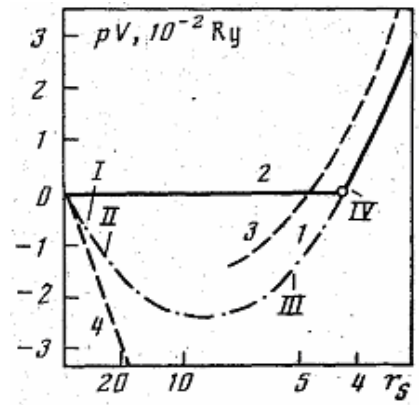

Fig. 3                               Fig.4

**Fig. 3**. Melting boundary in jellium model with rigid background: *1*) boundary in the classical region ($\Gamma \approx 178$ [22]); *2*) cold melting boundary ($r_s \approx 67$) [32]; *3*) estimation of [31].

**Fig. 4**. EOS of jellium model for $T = 0$: *1*) Pseudo-equation of state for OCP with incompressible background [32] (Monte-Carlo); *2*) Equation of state for OCP with uniformly compressible background; *3*) Hartree-Fock approximation ($pV$, Ry/particle = $1.47\, r_s^{-2} - 0.3...r_s^{-1}$); *4*) EOS of Wigner crystal ($pV \approx -0.6...r_s^{-1} +...$); *I* - melting ($r_s \approx 67 \pm 5$); *II* - transition from a ferromagnetic liquid to a paramagnetic one ($r_s \approx 26$); *III* - Liquid spinodal ($\partial p/\partial V)_T = 0$ ($r_s \approx 5.4$); *IV* - Binodal point ($p = 0$).

For $T \ll T_{cr}$, the spinodals corresponding to the high-density and low-density phases approach the lines $r_s$ = const and $\Gamma$ = const, whose parameters are given in the table.

The increase in $\Gamma$ with $Z$ on the gas spinodal (Table 1) is due to dilution of the charge system by the background. The shift to the right for this spinodal in Fig. 1 (and the same for the line $\Gamma$ = const = 178) is due to the choice of the quantity plotted along the ordinate for all $Z$: the background electron concentration, which is related to the ion concentration $n$ by $n_e = Zn$.

For $Z \to 100$, the ion subsystem melting curve does not intersect the gas-liquid spinodal. Nevertheless, there should be an intersection with the gas-liquid binodal for any $Z$, i.e., with the boundary to the two-phase region. The binodal parameters have not been calculated yet, but they can readily be estimated for $T \to 0$, since in that case the low-density almost ideal system of charges plus background coexists with the combination of a dense charge crystal in highly degenerate ideal electron gas. The properties of the dense phase in that case are almost independent of temperature, and the coexistence boundary corresponds to the condition

$$p \sim 0, \qquad \mu^{(gas)}(n, T) = \mu^{(condens)}(n) \qquad (5)$$

Therefore, the gas-phase density falls exponentially along the binodal for $T \to 0$ ($n_1 \sim \exp\{-\text{const}/T\}$), and for any $Z$ it intersects the charge subsystem melting curve $\Gamma = 178$ ($n \sim T^3$).

Figure 2 shows schematically the general form of the phase diagram for this OCP classical point-charge model with an ideal electron Fermi gas background.

**Jellium model**. In the theory of metals and in other applications, extensive use is made of Wigner's OCP model for electrons in a passive positive-charge compensating background (the so-called gel model). The form of the phase diagram is also substantially dependent on the assumptions about the background compressibility. If the background is incompressible, the only phase transition is crystallization, but there is a difference from the case given in Fig. 1 in that the growth in the zero-point electron oscillations causes the existence region for the crystalline phase to be additionally bounded on the high-density side [30] (Fig. 3).

At low temperatures ($kT \ll$ Ry), the boundary to the crystalline phase on the low-density side ($n_e \Lambda_e^3 \ll 1$) always also corresponds to $\Gamma = 178$ [22]. On the high-density side ($n_e \Lambda_e^3 \gg 1$) the available estimates of the melting parameters differ considerably [2, 32]. Approximate Monte Carlo calculations for $T = 0$ [32] predict melting for $r_s = a/a_0 = 67 \pm 5$ ($a_0$ is the Bohr radius, $a \equiv (3/4\pi n_e)^{1/3}$).

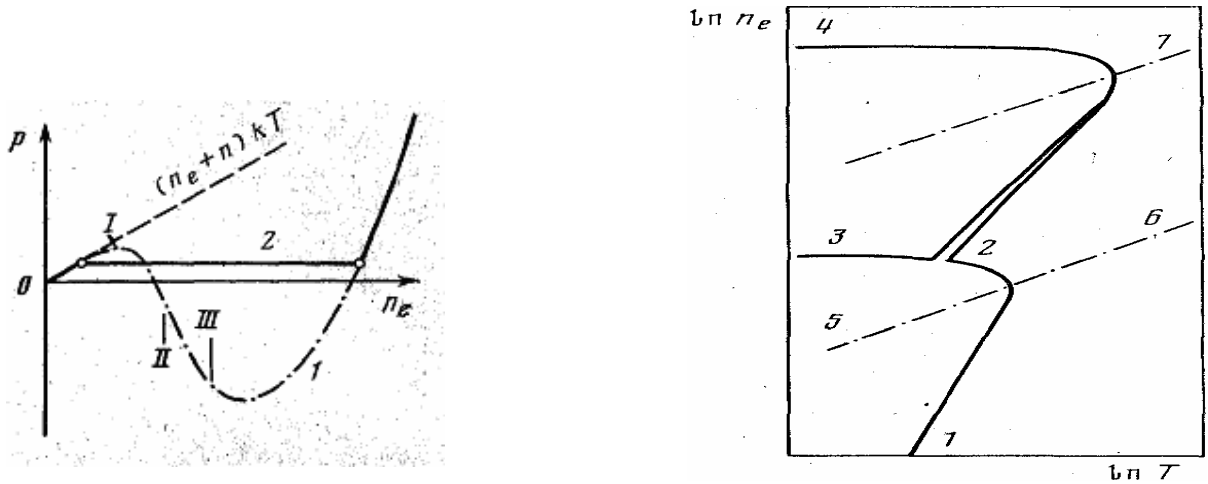

**Fig. 5.** Schematic form of the equation of state in the jellium model for $T \neq 0$ ($kT \ll$ Ry): *1* and *2* – see Fig.4; *I* - $(\partial p/\partial V)_T = 0$ ($\Gamma \approx 3$); *II* - crystallization ($\Gamma \approx 178$ [22]); *III* - melting point ($r_s \approx 67$ [32]).

**Fig. 6.** Schematic form of the global phase diagram for the Double-OCP model (superposition of OCP nuclei plus electron jellium): *1* - low density phase ("gas"); *2* - high density phase ("liquid"); *3* - nuclear crystal; *4* - boundary of 'cold' nuclear melting, $r_S \approx 67$; *5* – two-phase coexistence region; *6, 7* - degeneracy boundaries for electrons and nuclei correspondingly.

There is a resemblance to the classical OCP model in that transfer to a more realistic assumption as to nonzero background compressibility in the electron-gas model radically alters the form of the phase diagram because an additional phase transition occurs. There are in particular reasons to assume that the model has no space for crystallization, since, the corresponding values for the thermodynamic parameters fall entirely in the range where the system is thermodynamically unstable with respect to spontaneous volume change, i.e., in the two-phase gas-liquid coexistence region.

Such a suggestion has been made repeatedly in the literature for the case $T \approx 0$ [33, 34] on the basis of extrapolating approximations applicable for $r_s \ll 1$. Numerous approximate calculations have been made for the region $r_s \geq 1$, and these give more basis for the assertion (see review [2, 35]). If, in particular, we assume that the Monte Carlo calculations of [32] are exact, then the boundary to the Wigner crystallization ($r_s \approx 67$) lies far in the region where the electron OCP pressure and compressibility are negative (Figs. 4 and 5).

In the case $T \neq 0$, the available calculations on the equation of state for the gel model [36] are insufficient to define the boundaries of the Wigner crystallization and the above gas-liquid transition. An estimate based on Lindeman's criterion predicts a melting curve [31] lying in the region bounded by the inequalities $r_s \geq 67$ in Figs. 1 and 3. If we also assume that the boundary to the two-phase gas-liquid transition region in the gel model is close to the analogous boundary in the point-ion OCP model with $Z = 1$ and ideal electron Fermi gas background, then Fig. 1 shows that the assertion that Wigner crystallization is impossible is certainly met.

**Double-OCP model**. The at first sight arbitrary procedure for selecting the thermodynamic parameters of the background above can be given more content by using a variational principle from statistical mechanics. According to this, the free energy is minimal [37] for the system as defined for the case of an arbitrary non-equilibrium distribution $\rho_N$ with the true equilibrium distribution $(\rho_N)^0$:

$$F[\rho_N] \equiv \text{Sp}\{\rho_N(\mathbf{H}_N^* + kT\ln\rho_N)\} \geq F[\rho_N^0] \qquad (6)$$

$$(\rho_N)^0 \equiv \exp(-\mathbf{H}_N^*/kT)/\text{Sp}\{\exp(-\mathbf{H}_N^*/kT)\}$$

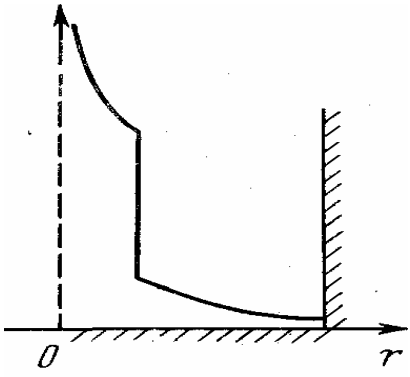 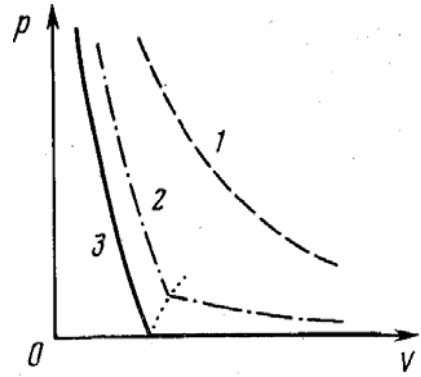

Рис. 7

Рис. 8

**Fig. 7.** Nucleus plus electron OCP in atomic cell (schematic form of the equilibrium electron density profile in the low-density limit for $T < T_{cr}^{OCP}$).

**Fig. 8.** Equation of State along isotherms for a system of independent mobile atomic cells filled with OCP of electrons: *1)* $T \gg T_{cr}^{OCP}$; *2)* $T < T_{cr}^{OCP}$; *3)* $T = 0$.

Simple application of (6) to the model considered here amounts to replacing the real distribution for the system of nuclei and electrons, $\rho_N$ by the product of two cofactors corresponding separately to the nuclear subsystem and the electron one:

$$\rho_{N(+)N(-)} = \rho_{N(+)}\rho_{N(-)} \qquad (7)$$

This operation denotes switching off the correlation between the two subsystems while completely retaining the correlations within each of them. The subsystems are then converted one for the other into compensating (passive) charge backgrounds of opposite signs. Therefore, the substitution of (7) and the minimization of (6) replace the real plasma by a superposition (combination) of two OCP models (nuclei and electrons), which are as it were embedded one in the another. The term Double-OCP model has been proposed for this combination [38].

The subsequent exclusion of all the dynamic correlations in one of the subsystems of (7) converts it to the corresponding ideal gas with a Hartree-Fock term for the interaction correction. We neglect the latter, which increases the free energy for Fermi particles, to arrive at one of the two OCP models discussed above: **a)** OCP for nuclei on the background of ideal electron Fermi-gas, and **b)** an electron OCP on the background of an ideal gas of positive nuclei (jellium). We can finally write [38]

$$F_{(+-)} \leq F_{(+)}^{(OCP)} + F_{(-)}^{(OCP)} \left\{ \begin{array}{l} \leq F_{(+)}^{(HF)} + F_{(-)}^{(OCP)} \leq F_{(+)}^{(ideal)} + F_{(-)}^{(OCP)} \\ \leq F_{(+)}^{(OCP)} + F_{(-)}^{(HF)} \leq F_{(+)}^{(OCP)} + F_{(-)}^{(ideal)} \end{array} \right\} \leq F_{(+)}^{(ideal)} + F_{(-)}^{(ideal)} \qquad (8)$$

Therefore, the Double-OCP model is the best according to (8) as regards the free energy it gives out of the set of simple Coulomb models in which one neglects the mutual screening for charges of opposite signs.

The phase diagram for the Double-OCP model (Fig. 6) has a form analogous to that of the OCP model for nuclei on a background of an ideal electron Fermi gas. The differences are the less the higher the nuclear charges number $Z$ in the OCP.

The above OCP models do not exhaust all varieties in which this phase transition should occur on going to a background compressed as a whole. In particular, this applies to the two-dimensional OCP model [39, 40], for example, an electron layer retained on a planar surface, and also to an OCP of classical Coulomb particles with additional hard-sphere repulsion at short distances [41] etc. [7].

This gas-liquid transition in a one-component plasma has so far attracted little attention, apart from [34], where estimates were made of the parameters of the coexisting low-density and high-density phases for an electron gas with an electrostatic background for $T \approx 0$. The behavior of the bounds as temperature increased was uncertain, as was whether they converge at all, but in [34] it was observed that the parameters required for Wigner crystallization are unattainable for $T \approx 0$. This agrees with the conclusions of [33] and in various other studies.

This lack of interest occurs in part because this transition in the OCP model lies outside the region where the model has applied value (i.e. $r_s \ll 1$); we give, however, examples of situations where the existence of this transition cannot be ignored.

### 1. Model of atomic cell with quasi-homogeneity approach for electrons.

In this approximation, a spherical atomic cell containing the nucleus at the center is filled with an electron 'liquid', which is described by a local equation of state coinciding with that for OCP of a macroscopic electron gas. In the simplest case, this is the equation of state for an ideal Fermi-gas (Thomas - Fermi model) [42]. If, on the other hand, one uses the equation of state of the interacting electron (jellium) model, the gas-liquid phase transition occurs at $T < T_{cr}^{OCP}$ not only when one calculates the equilibrium electron-liquid profile in the cell but also in the final electronic contribution in total equation of state. The cell electron-density profile $n_e(r)$ should show a discontinuity at sufficiently low densities, while there should be a kink on the isotherms in the equation of state. This is shown schematically in Figs. 7 and 8. The qualitatively similar picture should be valid also in the well-known Thomas-Fermi-Dirac approximation.

### 2. Non-equilibrium two-temperature plasma.

A situation can occur in a cold gas where an external source produces a non-equilibrium degree of ionization and at the same time a high electron temperature, although the ion temperature is low [43]. The electron subsystem is weakly non-ideal and does not participate in screening of the ions, thus providing conditions for using the OCP model. At a sufficiently low temperature, the ionic subsystem be comes highly non-ideal and can provide conditions for the phase transition discussed above. Estimates show that for example with $T_i \approx 100°K$ and $T_e \approx 1000°K$, the necessary ion and electron concentrations should be as follows: $n_e \sim 10^{17}$ cm$^{-3}$.

### 3. Ionization equilibrium in a plasma containing condensed-phase particles.

The case can occur [44] where the particles are negatively charged and collect dense screening clouds of positive ions around them. Such a plasma will be highly non-ideal ($\Gamma_D \gg 1$) for a certain combination of the initial parameters. Correct charge-distribution calculation then requires allowance for ion correlation. Qualitatively speaking, this correlation leads to additional ion-ion attraction and increased ion density at the surfaces of condensed particles, with charge higher than given by the standard calculation from the Poisson - Boltzmarin equation [44]. A simple way of correcting for this is to transfer from the Boltzmann (ideal-gas) dependence for the local charge density on the self-consistent pseudopotential in Poisson's equation to the non-ideal dependence corresponding to the local equation of state for classical ionic OCP with hard-sphere ion-ion repulsion at short distances. An inevitable consequence of that step is a discontinuous solution for the ion density (for $T < T_{cr}^{OCP}$), which is related to the gas - liquid phase transition in the ion OCP as discussed here.